\documentclass[12pt,a4paper]{article}
\usepackage{amsfonts,amssymb,amsmath,array,graphicx,tabularx,subfigure}

\title{Scattering of internal waves from small sea bottom inhomogeneities}
\author{A.~D.~Zakharenko}
\date{Il'ichev Pacific oceanological institute, Baltiyskay St. 43, Vladivostok, 41, 690041, Russia }

\begin{document}
\maketitle
\begin{abstract}
The problem of scattering of linear internal waves from small
compact sea bottom inhomogeneities is considered from the point of
view of mode-to-mode scattering. A simple formula for modal
conversion coefficients $C_{nm}$ , quantifying the amount of energy
that is scattered into the n-th mode from the incident field m-th
mode, is derived. In this formula the representation of
inhomogeneities by their expansions into the Fourier and
Fourier-Bessel series with respect to angular and radial coordinates
respectively are used.
        Results of calculations, performed in a simple model case, are presented.
        The obtained formula can be used for a formulation of the inverse problem, as it was done
in the acoustic case \cite{za1,za2}.

\end{abstract}

\section{Introduction}
The concept of mode-to-mode scattering was considered in the context
of the acoustic scattering from small compact irregularities of the
ocean floor by Wetton, Fawcett \cite{we-fa}. In their work some simple formulas
for modal conversion coefficients, quantifying the amount of energy
that is scattered from one normal mode of the sound field to
another, were derived. Recently new formulas for these coefficients
were obtained by Zakharenko \cite{za1} and applied to the inverse scattering
problem in the subsequent work \cite{za2}. This paper contains the detailed
derivation of such formulas in the case of scattering of linear
internal waves from small compact sea bottom inhomogeneities. Some
numerical examples are presented.

\section{Formulation and derivation of the main result}

We shall use the linearized equations for  inviscid, incompressible
stably stratified fluid, written for the harmonic dependence on the
time with the factor $e^{-i\omega t}$ in the form
\begin{equation}\label{1}
\begin{split}
-i\omega\rho_0u+\beta P_x=0&,  \\
-i\omega\rho_0v+\beta P_y=0&, \\
-i\omega\rho_0w+\beta P_z+\beta\rho_1=0&, \\
-i\omega\rho_1+w\rho_{0z}=0&,  \\
u_x+v_y+w_z=0&,
\end{split}
\end{equation}
where $x$, $y$, and $z$ are  the Cartesian co-ordinates with the
z-axis directed upward, $\rho_0=\rho_0(z)$ is he undisturbed
density, $\rho_1=\rho_1(x,y,z)$ is the perturbation of density due
to motion, $P$ is the pressure, and $u$, $v$ $w$ are the $x$, $y$
and $z$ components of velocity  respectively. The variables are
nondimensional, based on a length scale $\bar h$(a typical vertical
dimension), a time scale ${\bar N}^{-1}$ (where $\bar N$ is a
typical value of the Brunt-V\"ais\"ala frequency), and a density scale а
$\bar \rho$ (a typical value of the density). The parameter $\beta$
is $g/(\bar h {\bar N}^2)$, where $g$ is the gravity acceleration.

The boundary conditions for these equations are
\begin{equation} \label{2}
\begin{split}
 w =  0\quad  \mbox{at}&\quad z=0,\\
 w =  -uH_x-vH_y\quad \mbox{at}&\quad  z=-H,
 \end{split}
\end{equation}
where $H=H(x,y)$ is the bottom topography.

We introduce a small parameter $\epsilon$, and postulate that the
components of velocity and the pressure are represented in the form
$$
u  =  u_0+\epsilon u_1+\ldots \mbox{,\qquad}
v  =  v_0+\epsilon v_1+\ldots \mbox{,}
$$
$$
w  =  w_0+\epsilon w_1+\ldots \mbox{,\qquad}
P  = P_0+\epsilon P_1+\ldots
$$

We suppose also that the bottom topography is represented in the
form $H = h_0+\epsilon h_1$, where $h_0$ is constant and
$h_1=h_1(x,y)$ is a function of $x$, $y$ vanishing  outside the
bounded domain $\Omega$, which in the sequel is called a domain of
inhomogeneity.

Excluding from the system (\ref{1}) $\rho_1$ and substituting the
introduced expansions, we obtain

\begin{equation} \label{3}
\begin{split}
  -i\omega\rho_0(u_0+\epsilon u_1+\ldots)+ \beta(P_{0x}+\epsilon P_{1x}+\ldots)=0&,
   \\
  -i\omega\rho_0(v_0+\epsilon v_1+\ldots)+ \beta(P_{0y}+\epsilon P_{1y}+\ldots)=0&,
  \\
  (\omega^2\rho_0+\beta\rho_{0z})(w_0+\epsilon w_1+\ldots)+
i\omega\beta (P_{0z}+\epsilon P_{z1}+\ldots)=0&,
 \\
 (u_{0x}+\epsilon u_{1x}+\ldots)+ (v_{0y}+\epsilon v_{1y}+\ldots)
+w_{0z}+\epsilon w_{1z}+\ldots=0&,
\end{split}
\end{equation}
with the boundary conditions
\begin{equation}\label{4}
\begin{split}
 w_0+\epsilon w_1+\ldots =  0  \quad \mbox{at}&\quad z=0,  \\
 w_0+\epsilon w_1+\ldots =  -(u_0+\epsilon u_1+\ldots)
(h_{x0}+\epsilon h_{1x})\qquad \qquad      \\
 -\epsilon(v_0+\epsilon v_1+\ldots)(h_{1y}+\ldots)  \quad
\mbox{at}& \quad  z=-H.\\
\end{split}
\end{equation}

Separating terms in various orders of $\epsilon$, we obtain a
sequence of boundary problems.
\par
    At order  $O(\epsilon^{0})$ we have
\begin{equation} \label{5}
\begin{split}
 -i\omega\rho_0u_0+\beta P_{0x}=0,&  \\
 -i\omega\rho_0v_0+\beta P_{0y}=0, & \\
 (\omega^2\rho_0+\beta\rho_{0z})w_0+i\omega\beta P_{0z}=0,&  \\
 u_{0x}+v_{0y}+w_{0z}=0,&
\end{split}
\end{equation}
with the boundary conditions
\begin{eqnarray*}
& w_0 =  0\ & \mbox{at}\ z=0 \\
& w_0 =  0\ & \mbox{at}\ z=-h_0.
\end{eqnarray*}
Differentiating the third equation in (\ref{5}) twice with respect
to $x$ and twice with respect to $y$, summing obtained equations and
replacing $\beta(P_{0zxx}+P_{0zyy})$ by $-i\omega(\rho_0 w_{0z})_z$,
we obtain\sloppy
\begin{equation}
(\omega^2\rho_0+\beta\rho_{0z})(w_{0xx}+w_{0yy})+\omega^2(\rho_0
w_{0z})_z=0.  \label{6}
\end{equation}
We seek a solution to this equation in the form of the sum of normal
modes $w_0=e^{i(kx+ly)}\phi(z)$, where $\phi$ is the eigenfunction
of the spectral boundary problem
\begin{equation} \label{7}
\begin{split}
 -(\omega^2\rho_0+\beta\rho_{0z})(k^2+l^2)\phi+\omega^2(\rho_0\phi_z)_z=0,&  \\
 \phi(0)=\phi(-h_0)=0,&
\end{split}
\end{equation}
with the eigenvalue $\lambda=k^2+l^2$. It is well known that the
problem (\ref{7}) has countably many eigenvalues $\lambda_n$, which
are all positive. The corresponding real eigenfunctions $\phi_n$
we normalize by the condition
\begin{equation} \label{8}
 -\int_{-h_0}^{0}\left(\omega^2\rho_0+\beta\rho_{0z}\right)\phi^2\,dz=
\frac{\omega^2}{k^2+l^2}\int_{-h_0}^{0}\rho_0(\phi_z)^2\,dz=1.
\end{equation}
The eigenfunctions $\phi_n$ and $\phi_m$ with $n\ne m$ are also
orthogonal
\begin{equation} \label{9}
(\phi_n,\phi_m)=0
\end{equation}
with respect to the inner product
\begin{equation} \label{10}
(\phi,\psi)=-\int_{-h_0}^{0}\left(\omega^2\rho_0+\beta\rho_{0z}\right)\phi\psi\,dz
\end{equation}
In our scattering problem $w_0$ is the incident field, and we shall
calculate the main term of scattering field $w_1$, so we act in the
framework of the Born approximation.
\par
At the first order of $\epsilon$ we obtain the following system of
equations:
\begin{equation}\label{11}
\begin{split}
 -i\omega\rho_0u_1+\beta P_{1x}=0,& \\
 -i\omega\rho_0v_1+\beta P_{1y}=0,& \\
 (\omega^2\rho_0+\beta\rho_{0z})w_1+i\omega \beta P_{1z}=0,&  \\
 u_{1x}+v_{1y}+w_{1z}=0,&
\end{split}
\end{equation}
with the boundary conditions
\begin{equation} \label{12}
\begin{split}
w_1 =  0 \quad & \mbox{at}\quad z=0, \\
w_1 =  -u_0 h_{1x}-v_0h_{1y}\quad & \mbox{at}\quad z=-h_0 - \epsilon
h_1.
\end{split}
\end{equation}

So far as we are interesting in the connection of modal contents of
incident and scattering fields, we suppose that the incident field
consists of one mode $w_0=e^{i(k_nx+l_ny)}\phi_n(z)$. Reducing the
second boundary condition (12) to the boundary $z=-h_0$ with taking
into account the explicit form of $w_0$, we obtain the new boundary
condition for $w_1$ at the boundary $z=-h_0$:
\begin{equation} \label{13}
w_1 = \left(h_1 -\frac{ik_n}{k^2_n+l^2_n}h_{1x}
-\frac{il_n}{k^2_n+l^2_n} h_{1y}\right)e^{i(k_nx+l_ny)}\phi_{nz}.
\end{equation}
\par
Reducing the system (\ref{11}) in the same manner as it was done for
the system (\ref{5}), we obtain the equation for $w_1$:

\begin{equation}
(\omega^2\rho_0+\beta\rho_{0z})(w_{1xx}+w_{1yy})+\omega^2(\rho_0
w_{1z})_z=0. \label{14}
\end{equation}
We seek the scattering field in the form
$w_1=\sum_{m=1}^NC_{nm}(x,y)\phi_m$, the functions $C_{nm}(x,y)$ are
called the modal conversion coefficients. To obtain the equation for
$C_{nm}$ we substitute the postulated form of $w_1$ to the
(\ref{13}), multiplicate it by the function $\phi_m$ and integrate
from $-h_0$ to 0. Using the conditions of orthogonality and
normalization (\ref{8}), (\ref{9}) and the boundary condition
(\ref{13}), we finally obtain

\begin{equation} \label{15}
\frac{\partial^2}{\partial x^2}C_{nm}+\frac{\partial^2}{\partial
y^2}C_{nm}+ (k_m^2+l_m^2)C_{nm}=F,
\end{equation}
where
$$
F=\omega^2\rho_0\left(h_1 -
\frac{ik_n}{k^2_n+l^2_n}h_{1x} -\frac{il_n}{k^2_n+l^2_n}h_{1y}\right)
e^{i(k_nx+l_ny)}\phi_{nz}(-h_0)\phi_{mz}(-h_0).
$$
\par
Writing the solution to the equation (\ref{15}) as the convolution
of the fundamental solution (Green function) of the Helmholtz operator
$G =(-i/4)H_{0}^{(1)}(\sqrt{k_m^2+l_m^2}R)$  с with
the right-hand side $F$, we have
\begin{equation} \label{16}
C_{nm}(x_r,y_r) = -\frac{i}{4}\int\limits_x \int\limits_y F
H_{0}^{(1)}(\sqrt{k_m^2+l_m^2}R)\,dy\,dx,
\end{equation}
where $R=\sqrt{(x-x_r)^2+(y-y_r)^2}\,$ and by the index $r$ we
designate the point of registration of the field.
\par
Integrating by parts the terms containing $h_{1x}, h_{1y}$
and passing to the cylindrical coordinate system with the origin in
our domain of inhomogeneity and such that $k_n=\kappa_n$, $l_n=0$,
$x=r\cos\alpha$, $y=r\sin\alpha$, we obtain
\begin{equation} \label{17}
C_{nm}=-\frac{1}{4}\frac{\kappa_m}{\kappa_n}G\int\limits_0^{\infty}
\int\limits_0^{2\pi}h_1e^{i\kappa_nr\cos\alpha} \cos(\psi-\alpha_r)
H_{1}^{(1)}(\kappa_mR)r\,d\alpha dr,
\end{equation}
где $G=\omega^2\rho_0\phi_{nz}(-h_0)\phi_{mz}(-h_0)$,
$R=\sqrt{r^2+r_r^2-2rr_rcos(\alpha-\alpha_r)}$, $(r_r,\alpha_r)$ are
the polar coordinates of  the registration point,
$\tan(\psi)=r\sin(\alpha-\alpha_r)/(r_r-r\cos(\alpha-\alpha_r))$.
\par
Using the addition theorem for the Bessel functions we express
contained in (\ref{17}) $\cos\psi H_{1}^{(1)}(\kappa_mR)$ and
$\sin\psi H_{1}^{(1)}(\kappa_mR)$ in the form:
\begin{eqnarray*}
\left\{\vphantom{A}^{\cos(\psi)}_{\sin(\psi)}\right\}
H_{1}^{(1)}(\kappa_mR)=
\sum_{k=-\infty}^{\infty}H_{k+1}^{(1)}(\kappa_mr_r)J_k(\kappa_mr)
\left\{\vphantom{A}^{\cos k(\alpha-\alpha_r)}_{\sin
k(\alpha-\alpha_r)} \right\}\,.
\end{eqnarray*}
From now on we shall assume that the distance $r_r$ to the registration point is big enough
to replace the functions $H_{k+1}^{(1)}(\kappa_mr_r)$ by  their
asymptotics
$$
H_{k+1}^{(1)}(\kappa_mr_r)\approx\sqrt{2/(\pi\kappa_mr_r)}\exp{[i(\kappa_mr_r-(\pi/2)(k+1)-\pi/4)]}\,.
$$
Then, expanding $h_1(r,\alpha)$ as function of $\alpha$ in Fourier
series with the coefficients $\tilde h_{1\nu}(r)$, after integration
with respect to $\alpha$, we obtain
\par
\begin{equation} \label{18}
\begin{split}
 C_{nm}= & \frac{i\sqrt{2\pi}}{2}
\frac{\sqrt{\kappa_m}\exp(i\kappa_mr_r-i\pi/4)}{\kappa_n\sqrt{r_r}}G
\cos{\alpha_r}\sum_{\nu=-\infty}^{\infty}(i)^\nu
e^{-i\nu\alpha_0} \\ \\
& \qquad \times\sum_{k=-\infty}^{\infty}e^{-ik\alpha_r}
\int\limits_0^{\infty}\tilde h_{1\nu}(r)
J_k(\kappa_mr)J_{\nu+k}(\kappa_nr)r\,dr
\end{split}
\end{equation}
Changing the order of integration and summation we can achieve
further simplification by using the formula
$$
\sum_{k=-\infty}^{\infty}J_k(\kappa_mr)J_{\nu+k}(\kappa_nr)e^{-ik\alpha_r}=
J_{\nu}(\xi r)e^{-i\nu\theta},
$$
 where
$\xi=\sqrt{\kappa_m^2+\kappa_n^2-2\varkappa_m\varkappa_n\cos{\alpha_r}}$,
$\theta=\arctan\displaystyle\frac{\kappa_m\sin{\alpha_r}}
{\kappa_n-\kappa_m\cos{\alpha_r}}\,.$ We expand now the radial
coefficients $\tilde h_{1\nu}(r)$ on the segment $[0,L]$, where they
do not vanish, in the Fourier-Bessel series
$$ \tilde
h_{1\nu}(r)=\sum_{p=1}^{\infty}f_p^\nu
J_\nu\left(\frac{\gamma_p^\nu}{L}r\right)\,, $$ where $\gamma_p^\nu$
are the positive roots of  the function $J_\nu$,
$J_\nu(\gamma_p^\nu)=0$. Substituting this expansion in (\ref{18})
and taking into account that
$$
\int\limits_0^L J_\nu\left(\frac{\gamma_p^\nu}{L}r\right)
J_\nu(\xi r)r\,dr=\frac{-L^2\gamma_p^\nu J_\nu(\xi L)J'_\nu(\gamma_p^\nu)}
{{\gamma_p^\nu}^2-\xi^2 L^2},
$$
we obtain the final expression for modal conversion coefficients
\par
\begin{equation} \label{cfurbes}
\begin{split}
C_{nm}= & -\frac{iL^2\sqrt{2\pi}}{2}
\frac{\sqrt{\kappa_m}\exp(i\kappa_mr_r-i\pi/4)}{\kappa_n\sqrt{r_r}}G
\cos{\alpha_r} \\ \\
& \qquad\times\sum_{\nu=-\infty}^{\infty}(i)^\nu J_\nu(\xi L)
e^{-i\nu(\alpha_0+\theta)}
\sum_{p=1}^{\infty}f_p^\nu\frac{\gamma_p^\nu J'_\nu(\gamma_p^\nu)}
{{\gamma_p^\nu}^2-\xi^2 L^2}\,.
\end{split}
\end{equation}

\section{Numerical examples}

For a model example we choose $\rho=e^{-\lambda
z}$,$\beta=\lambda^{-1}$ and $H=1$ . Then the spectral boundary
problem is written in the form
\begin{equation*}
\begin{split}
 \omega^2\phi_{zz}-\omega^2\lambda\phi_z-\kappa^2(\omega^2-1)\phi=0,&  \\
 \phi(0)=0\,,\quad \phi(-1)=0.&
\end{split}
\end{equation*}
The eigenfunctions of such a problem are $\phi=Ae^{\lambda
z/2}\sin((l+1)\pi z)$ with the eigenvalues
$$
\kappa=\frac{\omega\sqrt{(l+1)^2\pi^2+\lambda^2/4}}{\sqrt{1-\omega^2}}\,.
$$
Here $A=\sqrt{2}/(\sqrt{1-\omega^2})$ by the condition (9). For the
calculations the value of parameter $\lambda$ was taken to be equal
to $0.003$ , which corresponds to the typical stratification in the
ocean shelf zones. The domain of inhomogeneity has the form of the
ellipse with the big and small radii $a$ and $b$  of which were
taken in proportion $a:b=2:1$, and in this region
$$
h_1(x,y)=0.05\sqrt{1-\frac{x^2}{a^2}-\frac{y^2}{b^2}}\,.
$$
In the figure are presented the results of calculations with
$\omega=0.5$ and the angle of incident field $\alpha_0=0$, conducted
for various wave sizes $\kappa a$ of the scatterer. We note that
according to the meaning of the small parameter $\epsilon$, in these
calculations $\epsilon=0.05$.
    For the presentation of results we use the scattering amplitude
$$
F_{nm}(\alpha_r)=\left(\frac{e^{i\kappa_m
r_r}}{\sqrt{r_r}}\right)^{-1}C_{nm}(\alpha_r)\,.
$$

\begin{figure}
\subfigure[$|F_{11}|$]{
\includegraphics[width=0.4\textwidth]{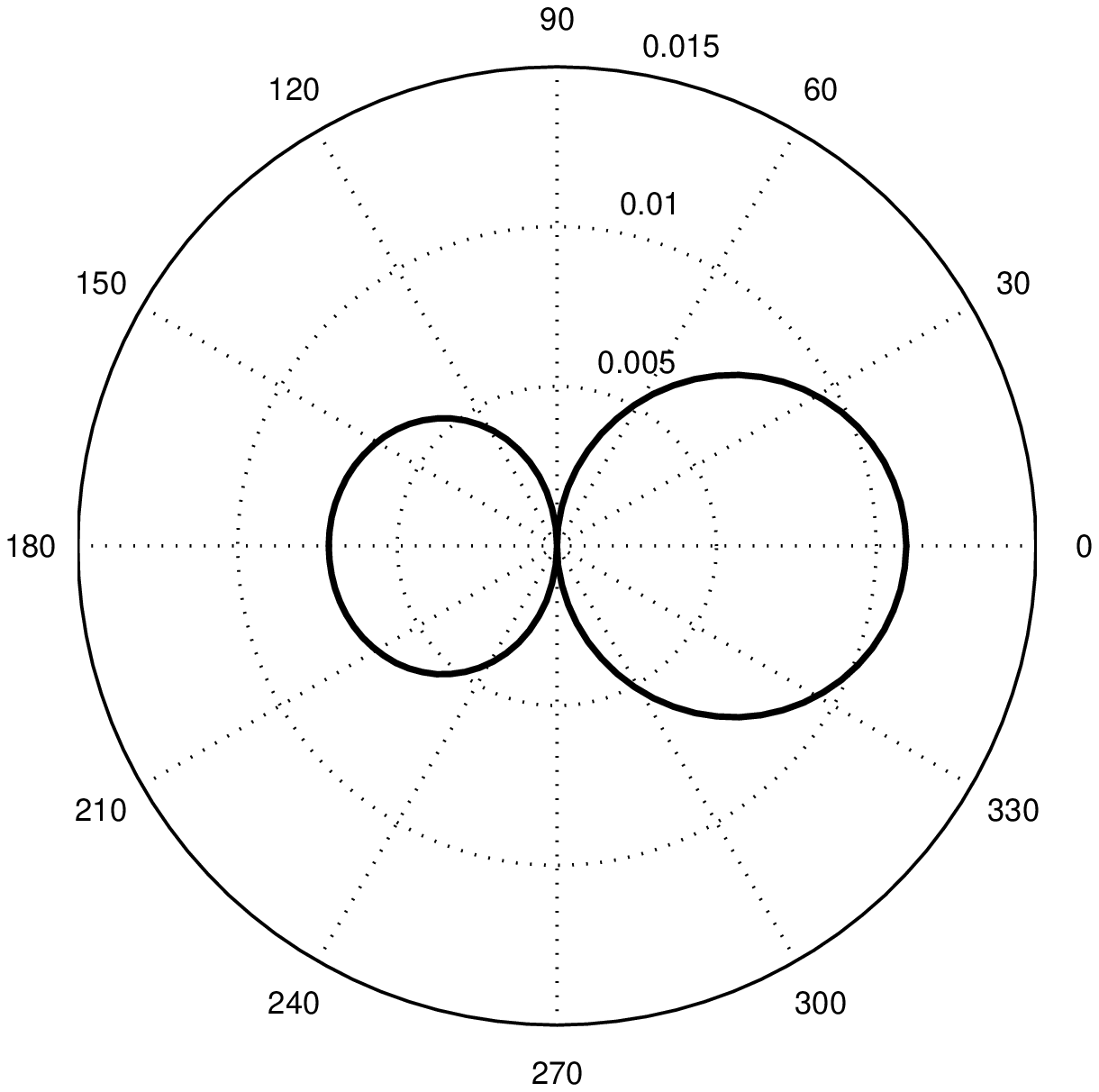}}
\subfigure[$|F_{12}|$]{
\includegraphics[width=0.4\textwidth]{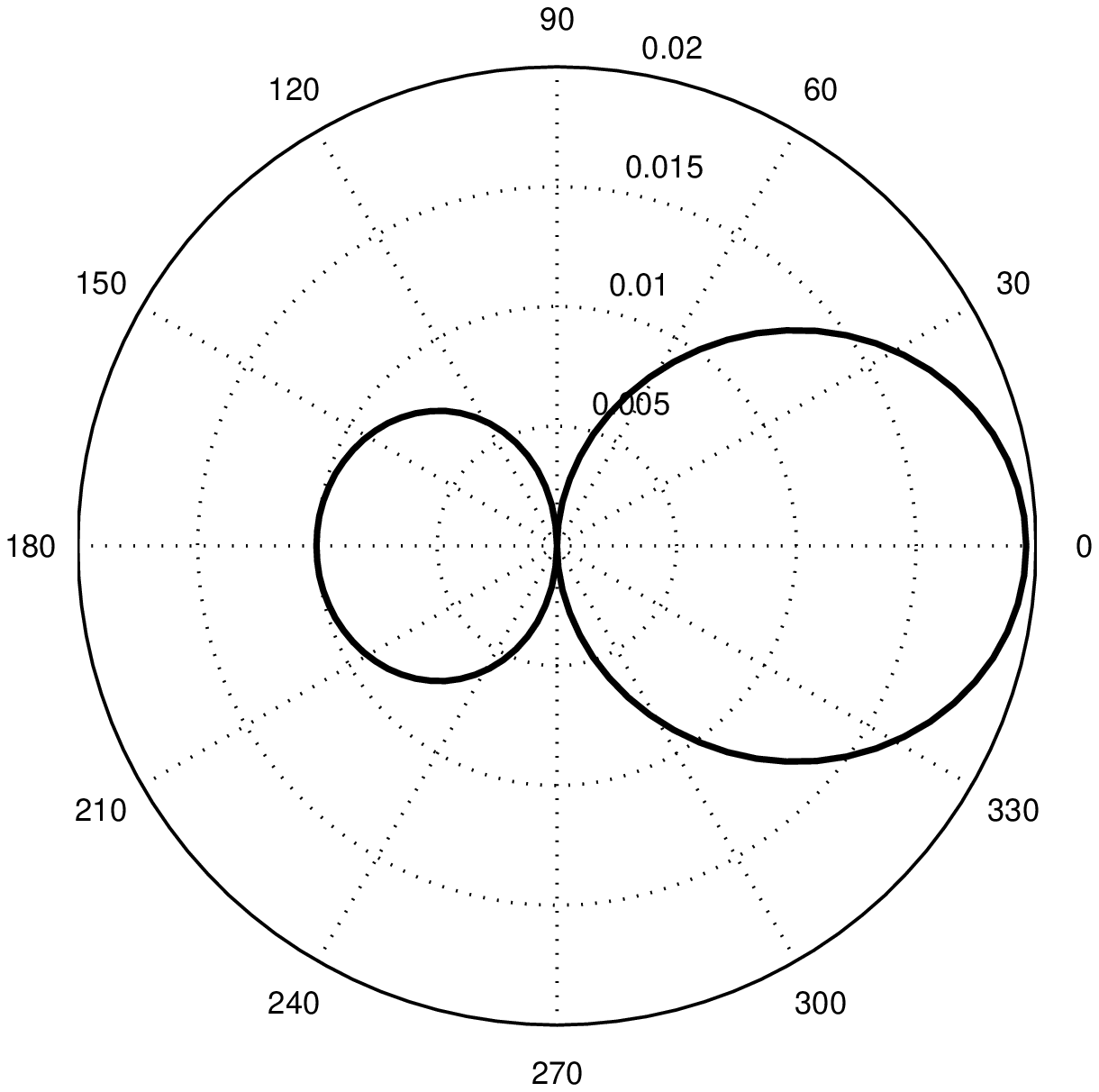}}
\subfigure[$|F_{11}|$]{
\includegraphics[width=0.4\textwidth]{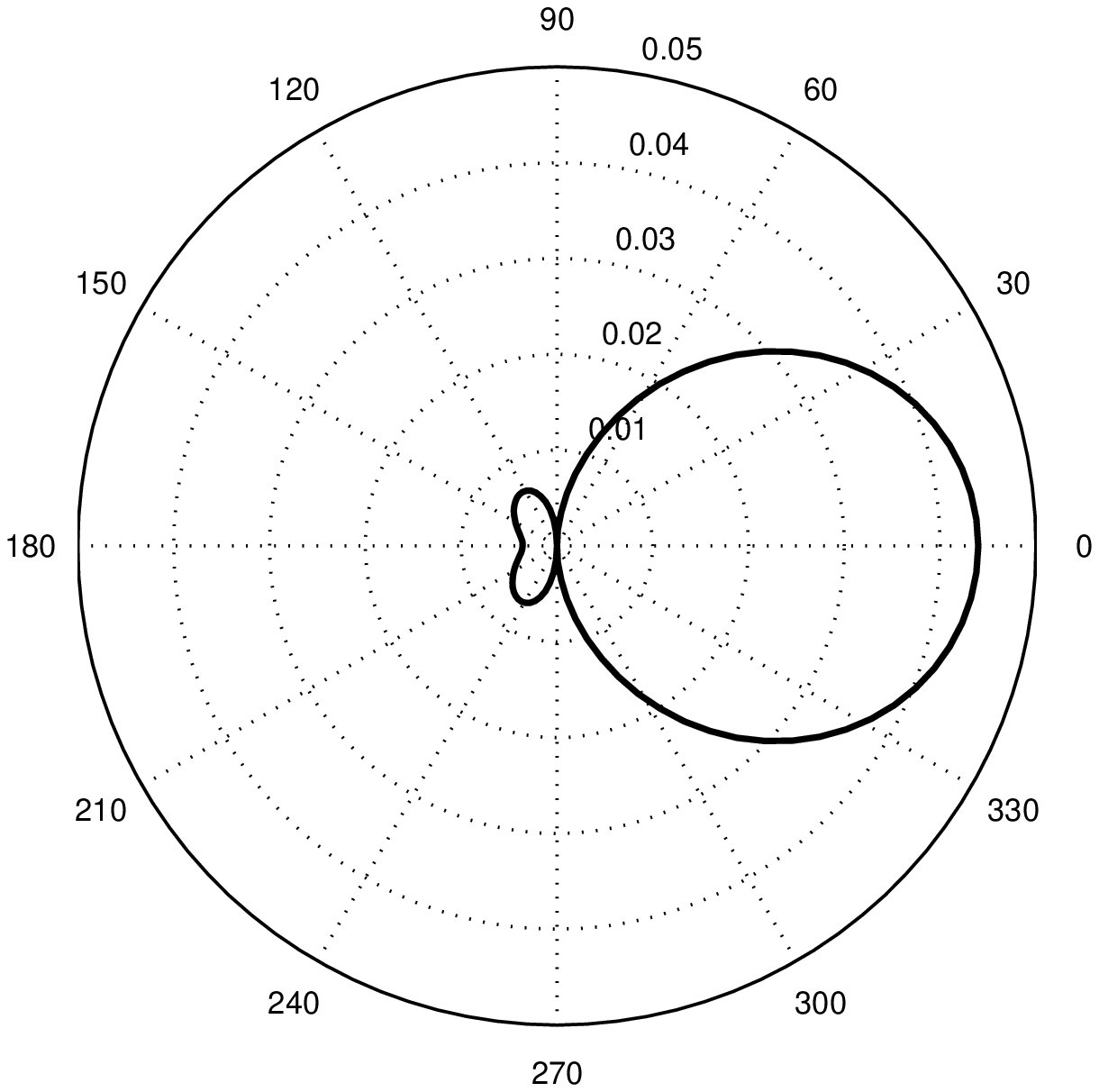}}
\subfigure[$|F_{12}|$]{
\includegraphics[width=0.4\textwidth]{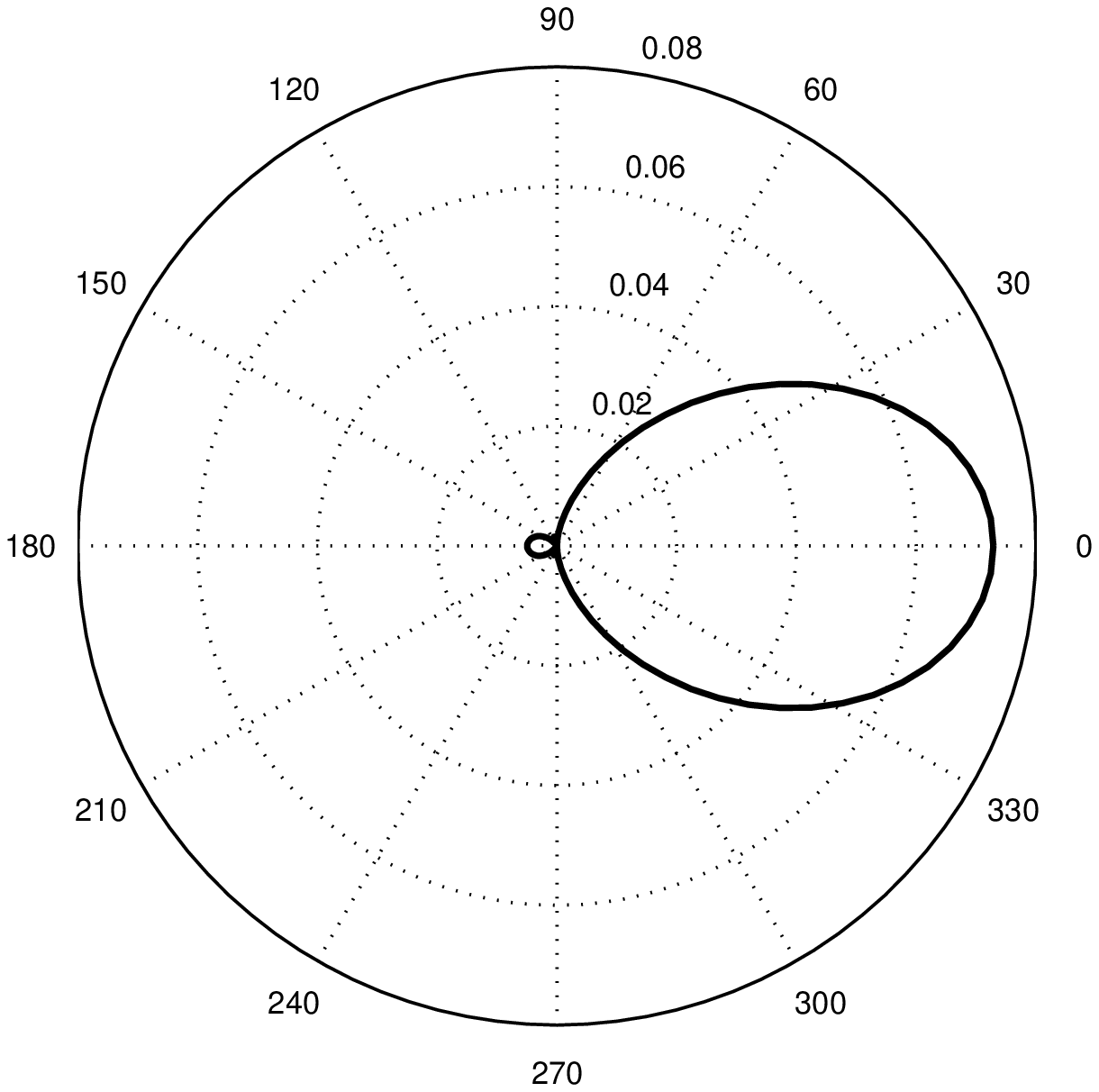}}
\subfigure[$|F_{11}|$]{
\includegraphics[width=0.4\textwidth]{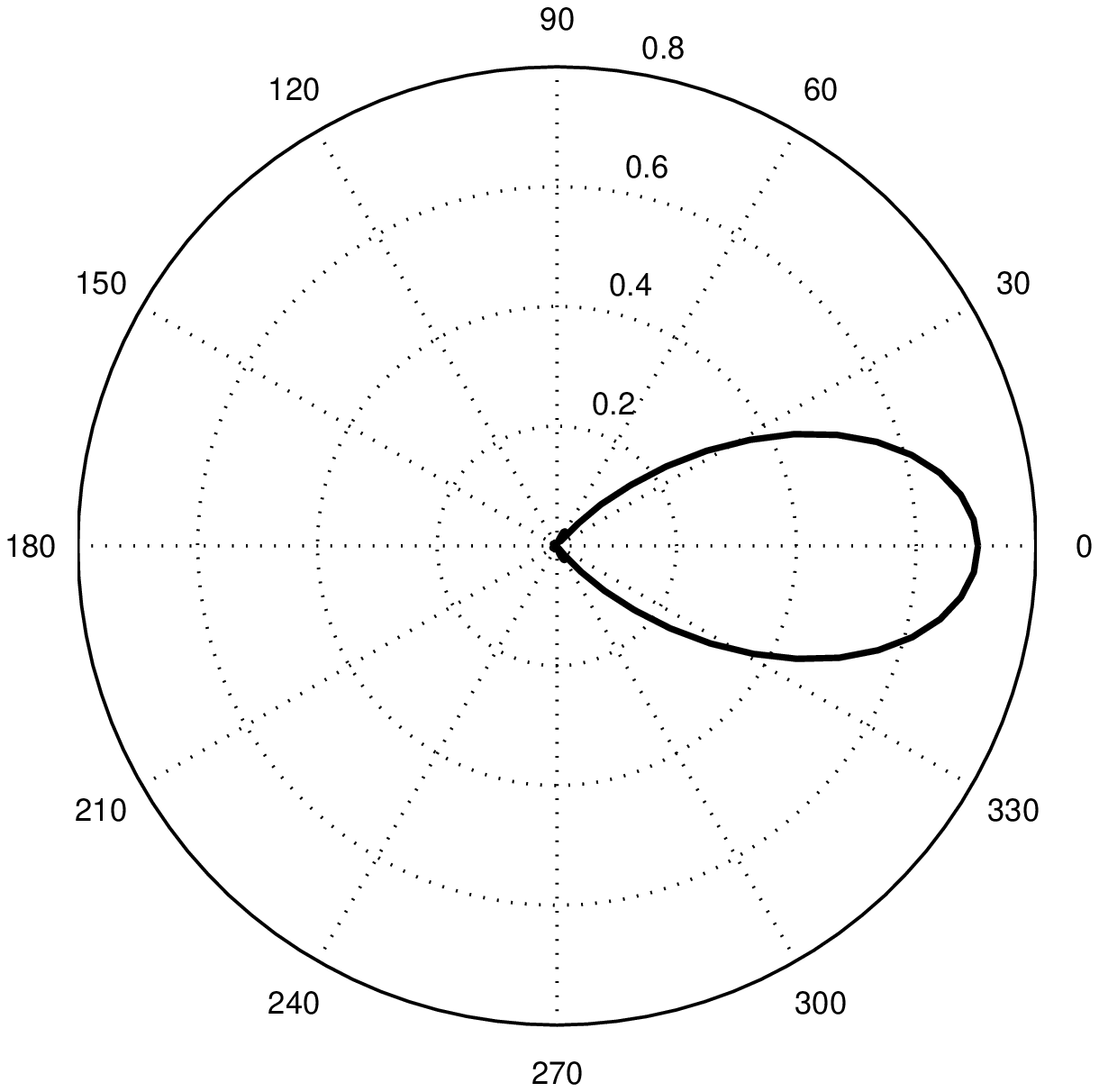}}
\hfill
\subfigure[$|F_{12}|$]{
\includegraphics[width=0.4\textwidth]{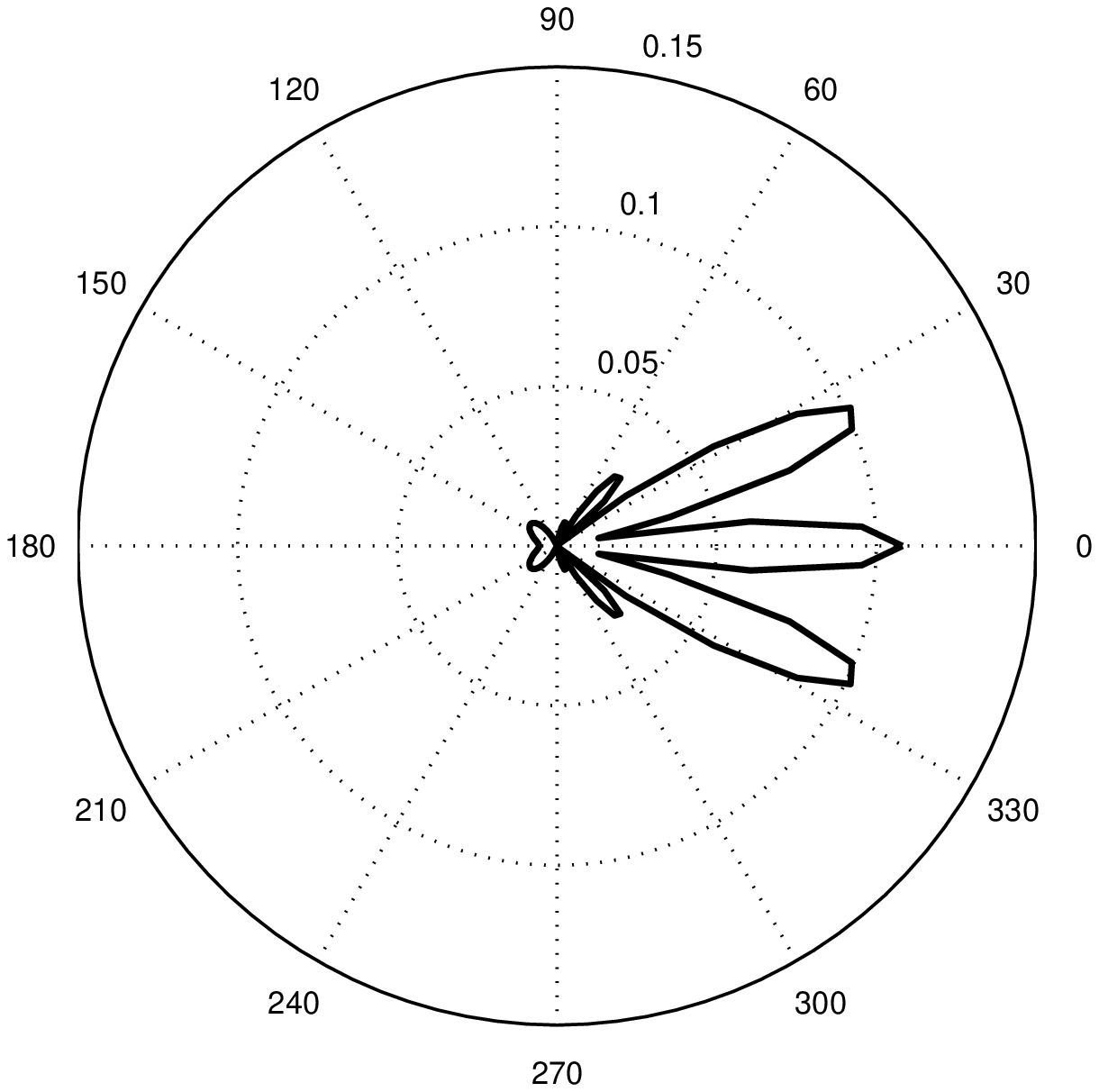}}
\caption{Absolute value of scattering amplitude: $\kappa a=1$ (a,b),
$\kappa a=2$ (c,d), $\kappa a=8$ (f,g)\label{fig1}}
\end{figure}

\end{document}